\documentclass[12pt]{article}

\usepackage{scicite}
\usepackage{graphicx}
\usepackage{hyperref}

\usepackage{times}

\newenvironment{sciabstract}{%
\begin{quote} \bf}
{\end{quote}}

\newcounter{lastnote}
\newenvironment{scilastnote}{%
\setcounter{lastnote}{\value{enumiv}}%
\addtocounter{lastnote}{+1}%
\begin{list}%
{\arabic{lastnote}.}
{\setlength{\leftmargin}{.22in}}
{\setlength{\labelsep}{.5em}}}
{\end{list}}

\title{Tuning order in cuprate superconductors}

\author
{Subir Sachdev$^{1\ast}$ and Shou-Cheng Zhang$^{2\dagger}$\\
\\
\normalsize{$^{1}$Department of Physics, Yale University,
P.O. Box 208120, New Haven, CT 06520-8120, USA}\\
\normalsize{$^{2}$Department of Physics, Stanford University,
Stanford, CA 94305, USA}\\
\\
\normalsize{$^\ast$URL:
\href{http://pantheon.yale.edu/~subir}{http://pantheon.yale.edu/{\~\,}subir}}\\
\normalsize{$^\dagger$E-mail:  sczhang@stanford.edu.} }

\date{{\em Science} {\bf 295}, 452 (2002)}

\begin{document}

% Double-space the manuscript.

\baselineskip24pt

% Make the title.

\maketitle

% Place your abstract within the special {sciabstract} environment.

\begin{sciabstract}
This article presents our perspective on STM measurements by
Hoffman {\em et al.} Science {\bf 295}, 466 (2002) of the vortex
lattice in Bi$_2$Sr$_2$CaCu$_2$O$_{8+\delta}$. We discuss
implications of these measurements for various theories of the
cuprate superconductors.
\end{sciabstract}

In 1986, superconductivity---the ability to transport electrical
current without significant resistance---was discovered in cuprate
compounds. These materials have fascinated physicists ever since,
in part because of the high critical temperatures ($T_c$'s) below
which superconductivity is present and the consequent promise of
technological applications. However, cuprate superconductivity
also raises fundamental questions about the collective quantum
properties of electrons that are confined to a lattice and
interact with each other (the ``correlated electrons'' problem).
Hoffman {\em et al.}\cite{seamus} have recently reported an
innovative scanning tunnelling microscopy (STM) study which should
help answer some of these questions.

All discussion of the cuprates begins with the compound
La$_2$CuO$_4$. Its valence electrons reside on certain $3d$
orbitals on the Cu ions which are arranged in layers. In each
layer, the Cu ions are located on the vertices of a square
lattice, and the ability of electrons to hop between successive
layers is strongly suppressed by the negligible inter-layer
overlap of the 3d orbitals. La$_2$CuO$_4$ is an insulator; its
inability  to transmit electrical current within a layer is a
result of the Coulomb repulsion between the electrons which
localizes them on the Cu sites. Moreover, it is known that the
spins of the electrons are oriented `up' and `down' in a
checkerboard pattern as shown in Fig. 1a: this quantum phase (or
state) is called an insulator with `N\'{e}el' or antiferromagnetic
order.

If we keep the material at zero temperature and tune another
parameter, such as the charge carrier concentration or the
magnetic field, we can explore different quantum states of the
system. For example, the properties of La$_2$CuO$_4$ change when
mobile charge carriers are introduced into the insulating N\'{e}el
state by chemical doping. In La$_{2-\delta}$Sr$_{\delta}$CuO$_4$,
a fraction $\delta$ of the electrons is removed from the square
lattice. The motion of the resulting holes is no longer impeded by
the Coulomb interactions and for $\delta>0.05$ the quantum state
(at zero K) of the electrons is a superconductor; this
superconductivity is present at all temperatures below $T_c$.

For $\delta>0.2$ and at low temperatures, the cuprates appear to
be qualitatively well described by the well-established
Bardeen-Cooper-Schrieffer (BCS) theory of superconductivity. In
this theory, the mobile electrons form pairs that condense into a
quantum state extending across the system. A gentle spatial
deformation of this state can set up a ``superflow'' of pairs,
leading to the phenomenon of superconductivity. However, the
internal wavefunction of the electron pairs has an unconventional
structure in the cuprates: the spins of the electrons are oriented
so that the total spin of the pair is zero, but their orbital
motion around each other is described by a wavefunction with
$d$-wave symmetry. In most low $T_c$ superconductors studied prior
to 1986, this wavefunction has an $s$-wave symmetry.

During the last decade, the debate has centered on the nature of
the quantum state of the cuprates at intermediate $\delta$---
between the well understood limits of the N\'{e}el insulator at
$\delta=0$ and the BCS superconductor at larger $\delta$. Many
candidate states have been proposed. A useful way of
characterizing them is in terms of different types of ``order'',
usually associated with breaking the symmetry of the electronic
ground state. For example, the N\'{e}el state breaks the symmetry
of spin rotations and lattice translations, and the superconductor
breaks the symmetry of charge conservation.

First, the order may be a spatial modulation of the local spin or
charge density\cite{zaanen,pnas,science,so5}. The simplest example
is the N\'{e}el state found in the insulator at $\delta=0$, and
shown in Fig. 1a. This state can be viewed as a wave in the spin
density, with a wavelength of two lattice spacings in the $x$ and
$y$ directions. At non-zero $\delta$, the wavelength of the spin
density wave changes; the orientation and period of this more
complex wave is described by a $\delta$-dependent wavevector ${\bf
K}$ (Fig. 1b). A charge density wave accompanies most such spin
density waves, with a wavevector of $2 {\bf K}$\cite{zachar}.
These spin and/or charge density waves are present at small
$\delta$, and  eventually vanish at one or more quantum critical
points leading to full restoration of invariance under spin
rotations and lattice translations.

In this picture, the order associated with superconductivity, and
with spin and charge densities should provide the foundation of a
theory of the cuprates at all $\delta$. At low $\delta$, the spin
density wave order dominates, resulting in a N\'{e}el state; at
high $\delta$, the order associated with superconductivity
dominates; at intermediate $\delta$, the two compete.

Second, the order may be associated with the fractionalization of
the electron\cite{pwa,senthil}. In certain theoretically proposed
quantum states, it is known that the electron falls apart into
independent elementary excitations, which carry its spin and
charge (such states need not break any symmetry). Experimental
tests for fractionalization have, however, not yielded a positive
signature so far \cite{moler,bonn}. A third set of
proposals\cite{lee,sf,varma} focuses on a rather unconventional
order linked with a spontaneous appearance of circulating
electrical currents, and an associated breaking of time-reversal
symmetry.

Given the distinct signatures of these proposals, one might expect
that experiments can resolve the situation quite easily. However,
the difficulty of smoothly varying the value of $\delta$ while
maintaining sample quality and avoiding extraneous chemical
effects has hampered progress. A recent set of
experiments\cite{gsb,bl1,bh,bk,bl2}, and especially in those
reported by Hoffman {\em et al.}\cite{seamus}, has led to a
breakthrough. These experiments show that it is possible to ``turn
a knob'' other than $\delta$ to tune the properties of the cuprate
superconductors. The ``knob'' is a magnetic field applied
perpendicular to the layers. Detailed dynamic and spatial
information on the evolution of the electron correlations as a
function of the applied field has been obtained. These data should
help solve the mystery of the cuprates.

Hoffman {\em et al.} studied a cuprate superconductor in an
applied magnetic field by a novel STM technology of atomically
registered spectroscopic mapping. The field induces vortices in
the superconducting order. Around each of the vortices is a
superflow of electron pairs. An innovative analysis of the large
amounts of STM data, with very high spatial and energy resolution,
enables Hoffman {\em et al.} to factor out the substantial noise
generated by chemical impurities introduced through doping, and
test directly for orders other than superconductivity.

Theoretical studies pointed out \cite{so5,arovas} that the
suppression of superconductivity in the vortex cores should induce
local magnetic order. This repulsion between the superconducting
and magnetic orders also appears in theories of magnetic quantum
phase transitions in the superconductor\cite{science,so5,csy}.
Combining these past works with insights gained from the neutron
scattering experiments by the group of Aeppli\cite{bl1,aeppli},
Demler {\em et al.}\cite{dsz} have pointed out that dynamic spin
density wave correlations (like those in Fig 1b) should be
enhanced in the regions of superflow which surround the much
smaller vortex cores. Static order in the associated charge
density wave has been proposed \cite{kwon}, in coexistence with
dynamic spin fluctuations and well-established superconductivity.
(see Fig 2).

Consistent with these expectations, the STM observations show a
clear modulation with a period of four lattice spacings in the
electron density of states around the vortices, in regions which
also display the characteristic signatures of electron pairing
associated with superconductivity. Moreover, the wavevector of
this ordering is $2 {\bf K}$, where ${\bf K}$ is the wavevector
for spin density wave ordering observed in neutron
scattering\cite{bl1} (albeit in a different cuprate
superconductor). The observed field dependencies of the neutron
scattering intensities \cite{bl1,bk,bl2} are also consistent with
theoretical expectations \cite{arovas,dsz}.

These observations are compelling evidence that the order
competing with superconductivity is the first of those discussed
above: a slight suppression of superconductivity reveals a
modulation in observables linked to the electron charge density.
This coexistence region between superconductivity and the
competing order should yield interesting new insights into the
fundamental properties of cuprates. Similar modulations should be
observable around other regions of the sample where the density
waves can be pinned, for example near impurities within the Cu
plane.

The scope for further studies using the magnetic field as a tuning
parameter is also wide. It should be possible to tune the cuprates
to the vicinity of quantum phase transition(s) associated with the
spin and charge ordering. Similar field-tuned studies can also be
carried out in other correlated electron systems, including the
electron-doped cuprates, organic superconductors, and
intermetallic compounds known as the heavy-fermion materials.

The next challenge will be to use our understanding of the low
temperature properties of the cuprate superconductors to formulate
a theory of competing orders above $T_c$. Here many mysteries
remain, particularly the microscopic origin of the ``pseudogap''
behavior, that is, the appearance of features characteristic of
energy gap of the superconducting state at temperatures well above
$T_c$.

\newpage

%\bibliography{scibib}

%\bibliographystyle{Science}

% Following is a new environment, {scilastnote}, that's defined in the
% preamble and that allows authors to add a reference at the end of the
% list that's not signaled in the text; such references are used in
% *Science* for acknowledgments of funding, help, etc.

\begin{scilastnote}
\item
We thank Ying Zhang for computations used in generating the second
figure.
\end{scilastnote}

\clearpage

\noindent {\bf Fig. 1.} {\bf Electron spin configurations on the
square lattice of Cu ions}. The arrows represent the direction and
magnitude of the average spin moment. The blue shading represents
the average electron charge density on each Cu site. ({\em a\/})
N\'{e}el state in the insulator at $\delta=0$. The spins oscillate
with a period of 2 lattice spacings in the $x$ and $y$ directions.
({\em b\/}) Density wave at a moderate value of $\delta$. A single
period of 8 lattice spacings is shown along the $x$ direction,
while the period along the $y$ direction remains at 2 lattice
spacings. Unlike Fig. 1a, the magnitude of the spin moment, and
not just its orientation, changes from site to site; we can also
expect \cite{zachar} a corresponding modulation of the charge
density on each site. The wavelength of the charge density wave is
half that of the spin density wave in both directions.

\noindent {\bf Fig. 2.} {\bf Magnetic field penetration of a
superconductor in a vortex state and the associated order.} The
superconducting order is suppressed at the cores of the vortices
(red dots). Superconducting currents (white loops) circulate
around the vortex cores. Experiment and theory discussed in the
text indicate that spin and charge orders depicted in Fig. 1b can
exist in the vortex state. The colored surface shows the envelope
of this order parameter, superimposed on the vortex lattice. This
type of order can be static or dynamically fluctuating depending
on the doping level and the magnetic field. The spacing between
the vortex cores is proportional to the inverse square root of the
applied magnetic field, and is typically about fifty times the
spacing of the lattice in Fig. 1.

%
% JPEG figures included below. Must use pdflatex to
% display them
%
\clearpage
\begin{figure}
\centerline{\includegraphics[width=5in]{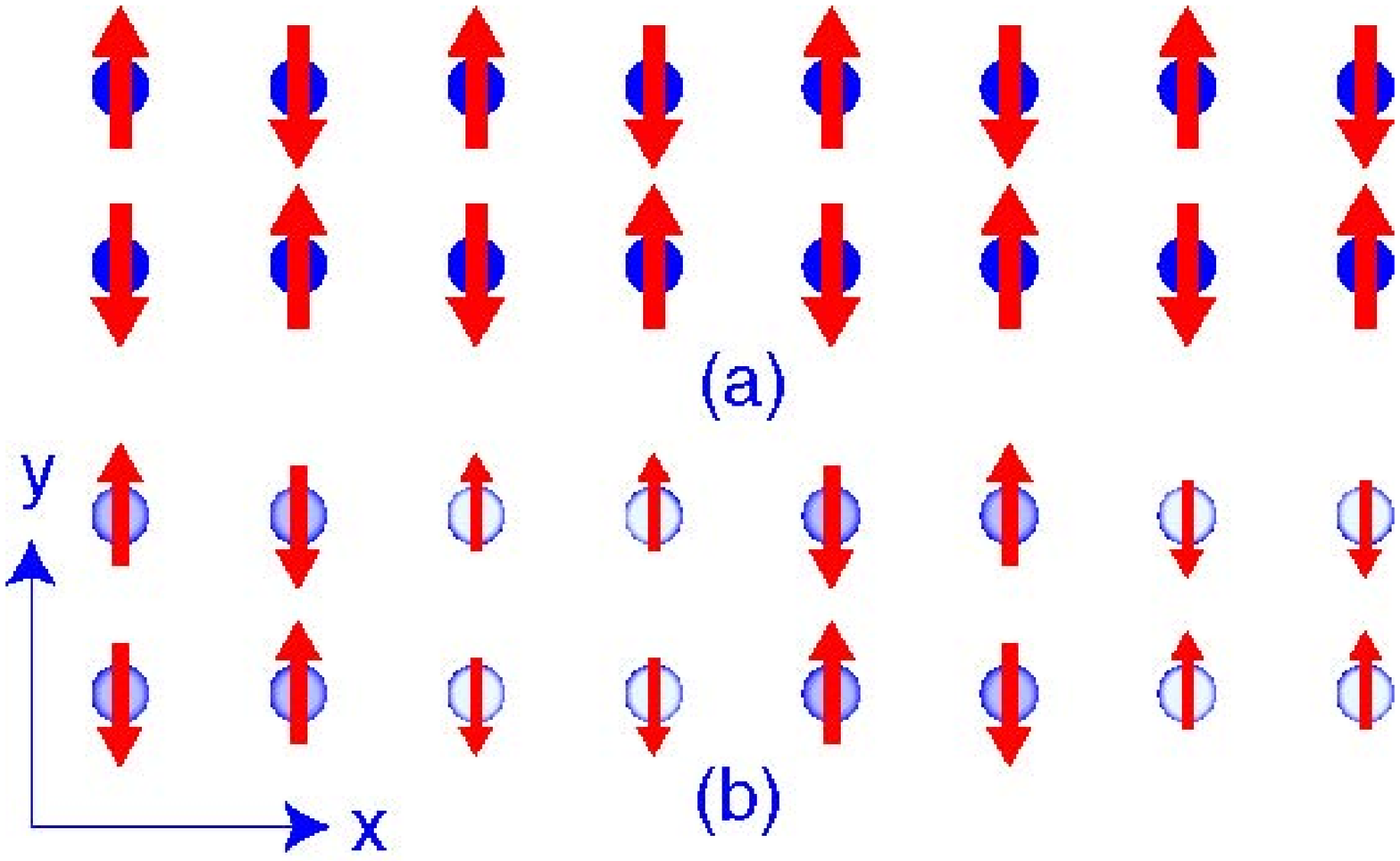}}
\vspace{0.5in}\caption{}
\end{figure}
\clearpage
\begin{figure}
\centerline{\includegraphics[width=7in]{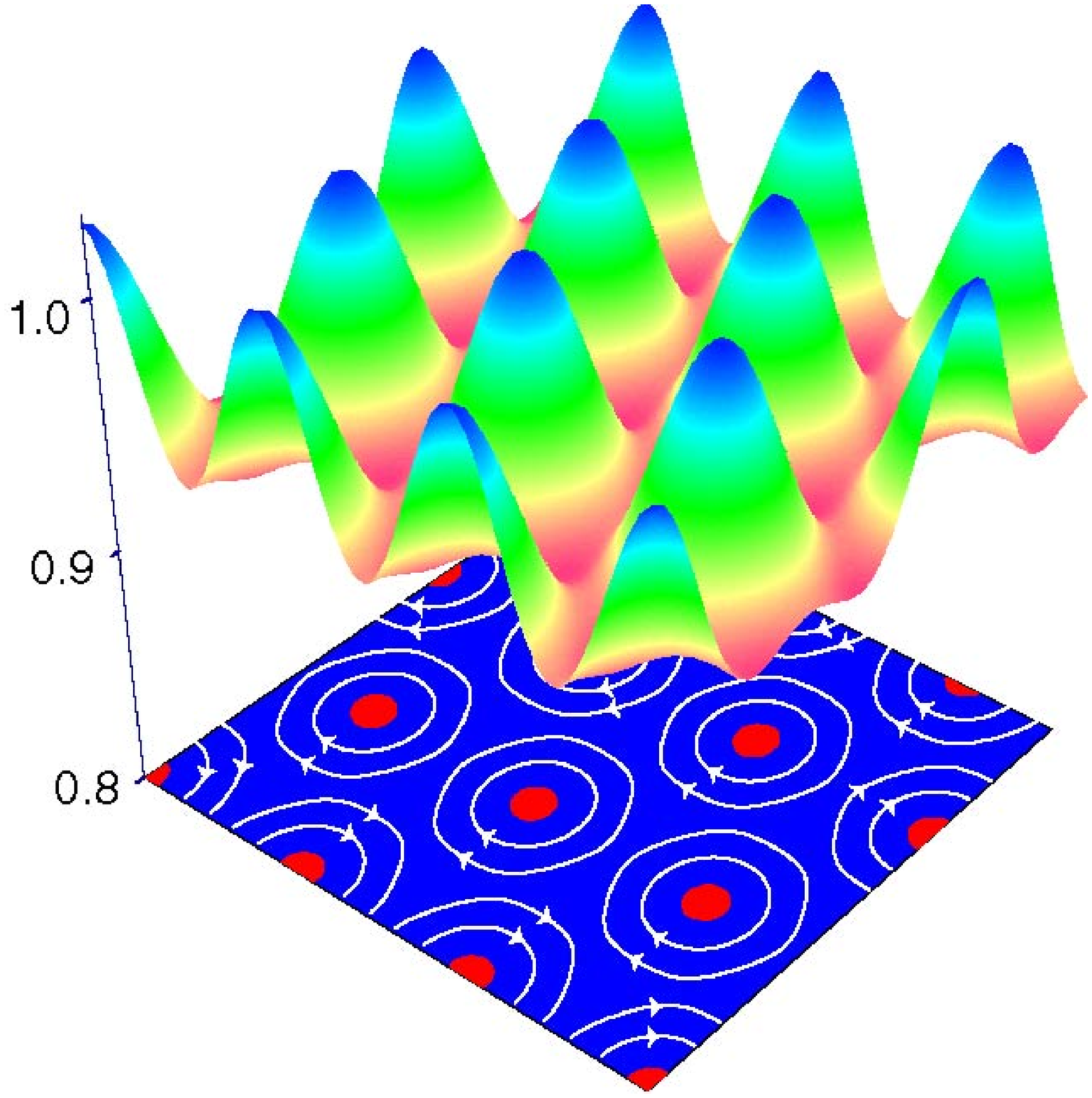}}
\vspace{0.5in}\caption{}
\end{figure}

\end{document}